# Nuclear Activity in Circumnuclear Ring Galaxies

**María P. Agüero[1], Rubén J. Díaz[2,3], Horacio Dottori[4]**

[1] Observatorio Astronómico and CONICET, Universidad Nacional de Córdoba, Laprida 854, 5000 Córdoba, Argentina.
Email: mpaguero@oac.uncor.edu
[2] ICATE, CONICET, Argentina
[3] Gemini Observatory, AURA, 950 N Cherry Av. Tucson, USA
Email: rdiaz@gemini.edu
[4] Instituto de Física, UFRGS, Av B. Gonçalves 9500, Porto Alegre, Brazil
Email: dottori@if.ufrgs.br

## Abstract

**We have analyzed the frequency and properties of the nuclear activity in a sample of galaxies with circumnuclear rings and spirals (CNRs). This sample was compared with a control sample of galaxies with very similar global properties but without circumnuclear rings. We discuss the relevance of the results in regard to the AGN feeding processes and present the following results: (i) bright companion galaxies seem not to be important for the appearance of CNRs, which appear to be more related to intrinsic properties of the host galaxies or to minor merger processes; (ii) the proportion of strong bars in galaxies with an AGN and a CNR is somewhat higher than the expected ratio of strongly barred AGN galaxies from the results of Ho and co-workers; (iii) the incidence of Seyfert activity coeval with CNRs is clearly larger than the rate expected from the morphological distribution of the host galaxies; (iv) the rate of Sy 2 to Sy 1 type galaxies with CNRs is about three times larger than the expected ratio for galaxies without CNRs and is opposite to that predicted by the geometric paradigm of the classical unified model for AGNs, although it does support the hypothesis that Sy 2 activity is linked to circumnuclear star formation. The possible selection effects of the sample are discussed, and we conclude that the detected trends are strong enough to justify high quality observations of as large as possible sets of galaxies with circumnuclear rings and their matched control samples.**

## 1. Introduction

The unified standard model for active galactic nuclei stands on a geometric paradigm, which implies that many of the observed properties among the different kinds of objects arise from the observer position and not from intrinsic properties of the host galaxy or their environment; in particular this statement is essential when explaining the differences among Seyfert types or among radio galaxy types. This strong constraint makes important any observational suggestion about systematic differences in the host galaxy properties that could not depend on the observer line of sight. A variety of statistical studies were carried out without conclusive results [1,2,3,4]. The confirmation of systematic differences in the intrinsic properties would introduce a new ingredient in the unified standard model, an evolution process from less energetic nuclei (Type 2 AGN) toward more energetic ones (Type 1 AGN), making the study of feeding processes crucial for understanding the nature of AGN classes.

Consequently, the connection between nuclear activity, star formation and infalling gas has received growing attention over the past ten years, with nuclear bars and circumnuclear disk instabilities being invoked as preferred mechanisms for removing angular momentum from the gaseous fuel [e.g. 5,6,7]. In order to finding any relation between AGN activity and environment, some structures present in the central regions of galaxies are particularly interesting; they are the star-forming circumnuclear rings and ring-like circumnuclear spirals (hereafter CNRs), which arise from the accumulation of infalling material at certain radius proving the active and effective removing of the gas angular momentum. Moreover, CNRs are morphological structures which are radially well differentiated from the active nucleus itself and other structures that can be associated to the active nuclei, like outflows or extended ionization regions. On the contrary, other circumnuclear star formation features as, for example, nuclear bars and hot spots can be confused at low resolution with AGN-related structures. Besides, CNRs represent a defined stage in the secular evolution of barred systems (e.g. [8]) making more probable the detection of any correlation with a defined AGN feeding stage.

The relationship between CNRs and the resonances (Inner Lindblad Resonance, ILR) that help to accumulate gas in these rings has been extensively discussed from both the theoretical [e.g. 9,10] and the observational [e.g. 11,12] standpoints. These resonances are a dynamical consequence of the bar perturbation but, although bars seem to contribute significantly to circumnuclear star formation, there are no robust evidence of a close relation between bars and nuclear "non-stellar" activity [13]. Even more some non-barred galaxies exhibit CNRs in their nuclear region, suggesting differences in time scale evolution of bars and CNRs.

In the present study we analyze a sample of nearby galaxies with CNRs and compare the properties of the host galaxy with that of a control sample, including the presence of AGNs and their type of activity. This work is organized as follow: In Section 2, we present the selected sample of galaxies with CNRs, we describe their properties which were considered to construct the control sample with similar properties than first one but without CNRs. In Section 3, we present the main results obtained from the comparison of both samples of galaxies. In Section 4, we carry out a discussion of the results presented in the previous section and offer the final remarks. The catalogue presented here contains the most comprehensive sample of circumnuclear rings extracted from the literature until to the year 2010. The previous version of this work is available at 2004astro.ph.9101.

## 2. CNR Galaxy Sample

### 2.1 Sample Description

We selected galaxies harboring CNRs in their central region. The CNR galaxy compilation was made based on images, color maps and radial profiles from literature up to 2006. The latter are required in order to be able to measure the CNR dimensions. The starting point in the search was the Catalogue of CNR galaxies from [14], catalog based on plate observations. They present 64 CNR galaxies of which we managed to get visual confirmation of 51 of them.

In spite of the rings are well defined structures, the classification of inner or circumnuclear rings is difficult when the global bar is not detected. For that reason we define some selection criteria in order to avoid misclassification of the observed ring. They are:

(i)     The diameter ratio of CNR respect to the host galaxy should be less than 0.2.
(ii)    CNR radius between 100pc and 2 Kpc. These restrictions were imposed based on the typical Inner Lindblad Resonances, limited for the Nuclear Lindblad Resonance and the inner ring at small and larger radii respectively.

A third criterion was added from the observational point of view. We are looking for well resolved features. Additionally, we are considering a statistical study of CNR galaxies, so we are interested in CNRs that can be observed from ground-based observations and that could be detected in massive surveys. Therefore, we selected CNRs with:

(iii)   Apparent CNR radius larger than 1.5 arcsec (Distance up to 275 Mpc).

ESO 565-G11 was exhaustively analyzed by [15] who determine that this galaxy harbors a CNR exceptionally large with a radius of 2.6Kpc. In spite of the CNR radius is out of the selection criterion, the circumnuclear classification of this ring is certain, and then this galaxy was included in the sample. ESO 138-G1 is a E/S0 galaxy with no apparent bar, which present a relative radius of 0.23, just above the impose limit of 0.20, but the absolute radius is 1.25 Kpc, well below of the upper radius limit, whereby ESO 565-11 was incorporated anyway.

Some galaxies were included in the present sample even without a clear image of the CNR, only if two or more authors claim the CNR existence, publishing the CNR geometrical parameters. In these cases, the CNR radius was assumed to be the mean of the cited values.

In Table 1, we list the 94 selected CNR galaxies. More than 30 galaxies are good candidates to have CNRs, however some of them do not fully satisfy our selection criteria and some others require detailed observations to confirm the presence of the ring.

The CNRs of the sample were classified depending on their evolutionary stage. The CNRs observed exclusively in the infrared bands or observed in abortion in optical bands were classified as Dust Rings; those observed in emission in optical bands were classified as HII Rings and, finally, the ringed structures only detected in color maps or using the imaging technique of "unsharp masking" were classified as Stellar Rings. The galaxy sample contains 73 HII Rings, 12 Dust Rings and 9 Stellar Rings. Galaxy diameter at 25mag in B-band adopted from NED (NASA/IPAC Extragalactic Database), the measured CNR radius and the ring-class are included in Table 1. The bibliography references of the source of CNR detection are presented in the table.

In Figure 1, we present the absolute and relative radius distribution of the sample. The mean radius is 800 pc (median of 700 pc). The 91% of the objects are between 200 pc and 1.85 Kpc, reinforcing the election of radius limits adopted in the sample selection criteria. Additionally, the upper limit of 20% in relative size is in the tail of the distribution, warranting the loss of a few objects because of it.

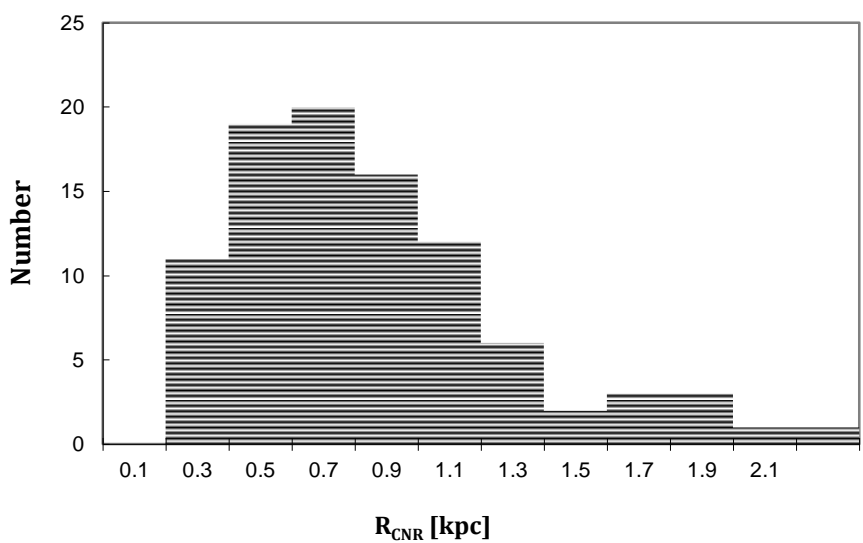

**Figure 1.** Distribution of Circumnuclear Ring Radius for 94 CNR galaxy sample.

The morphological type, bar classification, and nuclear activity for the selected galaxies are listed in Table 2. The first two characteristics were obtained from [16] and, for galaxies that were not catalogued by [16], from NED. The nuclear activity classification was based on the catalog of [17] who classify galaxies with strong nuclear activity, and we use [18] for classification of nuclei with lower activity. For the objects common in both catalog, the activity type was almost coincident. This suggests that the relative distribution of Seyfert types is not strongly dependent on the two classification sources used here. This nuclear activity classification is in agreement with that assigned by NED in all cases except for NGC 5850 which is classified as LINER in [18] but has no nuclear activity classification in NED.

The morphological distribution of the sample is shown in Figure 2a. As can be seen, the Sb and Sbc are the uppermost morphological types. In order to determine its significance, we compare this distribution with that expected from [16] (sample of 2367 nearby galaxies), and with that expected from [18] (sample of 486 spiral galaxies designed for a central region study). In the both reference distribution, the peak is observed in Sc galaxies. In figure 2b, we compare the CNR galaxy morphological distribution with [16] expected distribution for 94 objects. It is notorious, not only the Sb galaxies excess but also the overage of Sa type ($N^{1/2} = 6,2$; 5.7, for Sb and Sa galaxies respectively, with an excess of 10 objects respect to Tully expectation in both morphological types).

CNRs are thought to be related with the ILRs which are dynamically generated by the bar perturbation. However, some CNR galaxies are un-bared galaxies. In Figure 3, we display the distribution of bar intensity classification for CNR galaxies in comparison with the expected bar distribution from [16]. While the CNR galaxy sample presents, as expected, an excess of bared galaxies, the number of strong bars is the expected one. Indeed, the weak bared galaxies are out of the expectation respect of the Tully sample (16 objects above the expectation value, $N^{1/2} = 6,4$).

## 2.2 Nuclear Activity

The nuclear activity distribution of the sample is shown in Figure 4. The intermediate class of Seyfert galaxies, Sy1.2 and Sy1.5, were included in Sy1 class, while the intermediate class, Sy1.8 and Sy1.9 were added to the Sy2 class, as well as the Seyfert galaxies classified as Sy1 only in polarized light because this specific technique studies are available only for few galaxies, being classified as Sy2 in classic spectroscopic studies. The Sy2 galaxies are predominant in the CNR galaxy sample, even more frequent than nuclear star-forming galaxies. [19] established the frequency of Seyfert galaxies per morphological type for a sample of 1246 galaxies, whose nuclear activity classification was based on [20] catalog (Frequency of Sy1-Sy2: 2%-0% in E/S0; 6.5%-6.5% in Sa and Sab; 3.2%-3.9% in Sb and Sbc; 0.5%-0.8% in Sc). Considering the morphological distribution of the CNR galaxy sample shown in Figure 2, we determine the expected number of Sy1 and Sy2 galaxies in the sample, resulting in 3 Sy1 and 3 Sy2 galaxies. These expected values are very low respect to the observed ones of 6 Sy1 and 16 Sy2 galaxies. Thanks to the observational improvements, the number of discoveries of weak active nuclei was increasing respect to the active nuclei of higher activity. For that reason, we make a new comparison with a searching study of low activity in 486 nearby galactic nuclei, as is the case of [18] (Frequency of Sy1-Sy2: 0%-7% in E; 2%-7% in S0; 7%-12% in S0/Sab; 2%-12.6% in Sb and Sbc; 1%-4.5% in Sc and Scd; 0%-5% in Sd/Sm). Worth mentioning that this study was carried out with extremely detailed spectroscopic analysis looking for weak nuclear activity; however the CNR galaxy sample was not analyzed as thoroughly as that of [18]. Nonetheless, the expectation from [18] sample is 3 Sy1 and 9 Sy2 galaxies, still below the observed values. Considering the very low number of active nuclei in the CNR galaxy sample, we design another test trying to determine if the excess of active nuclei is significant. We build a comparison galaxy sample as a control sample.

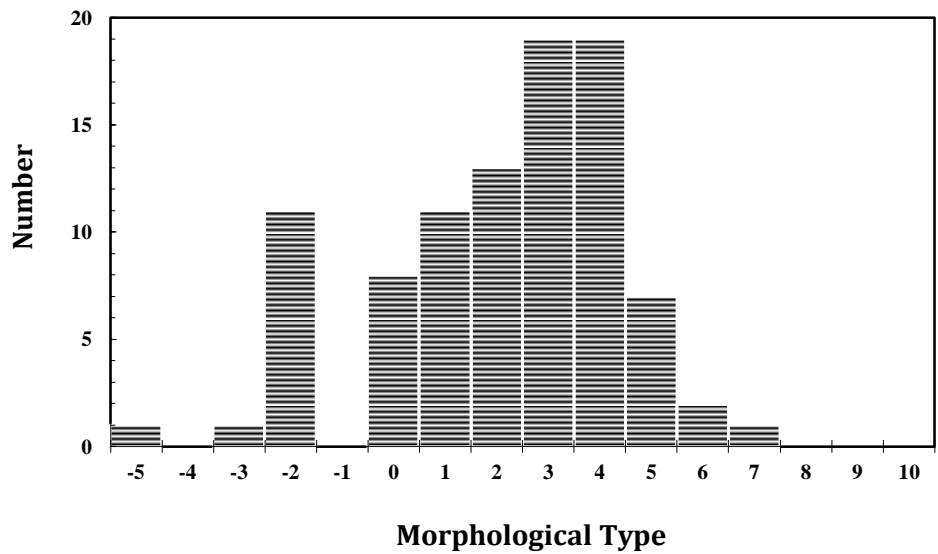

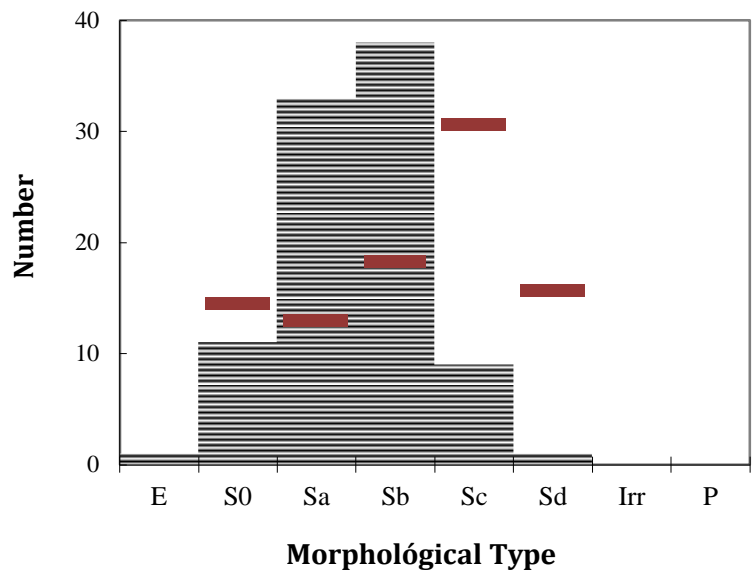

**Figure 2.** Distribution of Morphological Type from Tully (1998) for the 94 CNR galaxy sample. (Bottom) Comparison between the observed morphological distribution with that expected for a random sample of 94 galaxies from Tully (1998) Catalogue distribution.

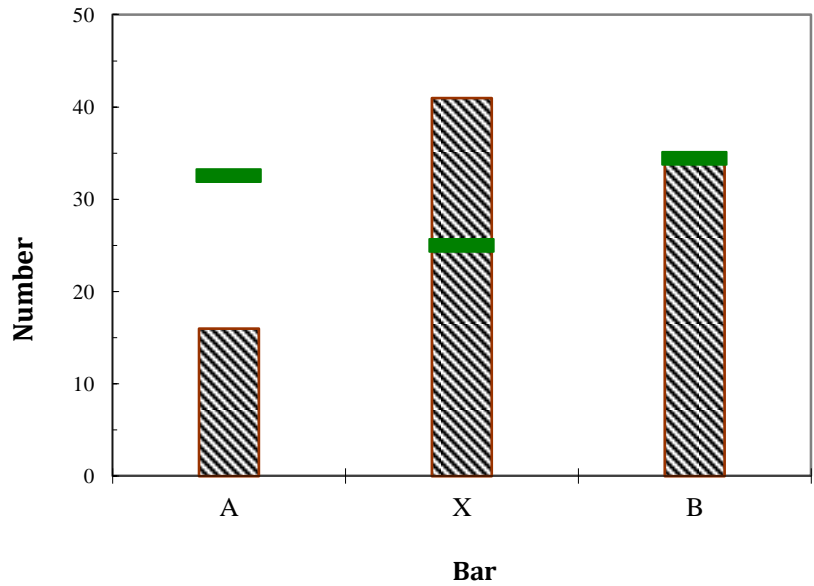

**Figure 3.** Distribution of bar classification for the 94 CNR galaxy sample. Marks indicate the expected distribution for a random sample of 94 galaxies from Tully (1998) Catalogue distribution.

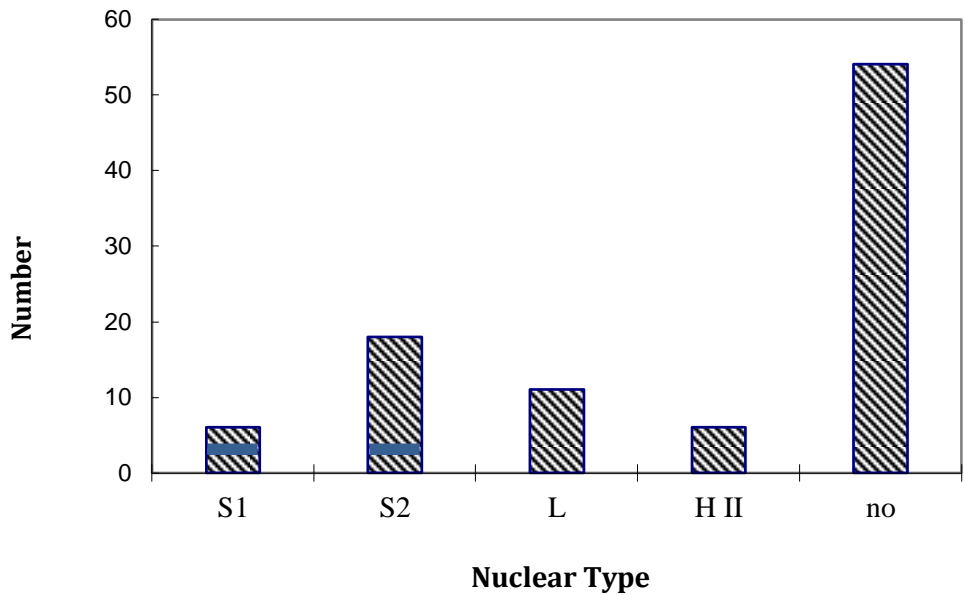

**Figure 4.** Distribution of nuclear classification for the 94 CNR galaxy sample. "S1" class includes Sy1 galaxies, Sy1.2, and Sy1.5; "S2" class includes Sy2 galaxies, Sy1.8, Sy1.9 and galaxies classified as Sy1 only in polarized light (Sy1i, Sy1h); "L" class includes LINER galaxies; "HII" class includes galaxies with star formation in the nucleus; "no" includes galaxies with stellar nuclei. Mrks indicate the expected Seyfert galaxy numbers from the Woltjer (1990) active galaxy frequency distribution considering the morphological distribution of the CNR galaxy sample.

## 3. Results

### 3.1 Comparison Sample:

We are interested in study the incidence of nuclear activity in the CNR galaxy sample. For that reason we excluded seven objects of the CNR galaxy sample because their CNRs were detected in studies of active galactic nuclei, so their active galaxy condition was decisive in the discovery of the CNR, conditioning the CNR and nuclear activity relationship. In order to avoid differences in the nuclear activity classification or global properties of CNR and comparison galaxies, we restricted the sample to those galaxies catalogued in [16] and we use exclusively the [17] catalog for nuclear activity classification.

The subsample of CNR galaxies has 67 objects. For the assignment of the comparison galaxy we consider the following aspects:

Observationally, the probability to detect a CNR depends on the inclination of the galaxy and the apparent diameter. Therein, we want the same CNR detection probability that the hitherto known CNR galaxies. Consequently, we impose the following conditions for comparison galaxies:

(i)      The departure in inclination from the CNR galaxy must be $\Delta i < 10°$;
(ii)     The difference in corrected apparent sizes must be $\Delta D_{25} < 0.2 \, D_{25}^{\, CNR}$

Physically, we are interested in comparison galaxies with similar global properties than the CNR galaxies in order to guarantee the same probability to have an active nucleus. Therefore, we designated the following criteria:

(iii)    The departure in morphological type numerical code from the CNR galaxy must be $\Delta T \leq 1$.
(iv)     The departure in B absolute magnitude from the CNR galaxy must be $\Delta M_B < 0.3$;
(v)      The difference in projected real sizes must be $\Delta R_{25}$ (kpc) $< 0.4 \, R_{25}^{CNR}$ (kpc);

We define a selection criteria index which includes the five selection criteria with the same weight each one. The galaxy in the [16] catalog that has the lowest selection index was assigned as the comparison galaxy for each CNR galaxy. The selected comparison galaxy was examined in the literature in order to guarantee that there is no indication of the presence of a CNR. In Table 3, we list the selection properties for CNR and comparison galaxies. The mean differences between both samples are (-4° ± 10°) in inclination; (0.8' ± 1.7') in apparent diameter; (0.0 ± 0.2) mag in B-band absolute magnitude; (0 ± 1) in morphological type; and (-0.9 ± 4.1) Kpc in absolute diameter.

The intensity bar distribution for both samples is shown in Figure 5. While the tendency is the same than the observed in the full sample (an excess of weak bars in CNR galaxies respect to the comparison galaxies), this result is not conclusive due to the low number of objects.

In the Figure 6, the incidence of active nuclei is plotted for both samples. It can observe more galaxies with active and star-forming nuclei in the CNR subsample than in the comparison one. The expectation for CNR and Sy2 galaxies obtained from [19] is of 2 Sy1 and 2 Sy2 for both samples (they have the same morphological distribution). Therefore, the number of Sy1 galaxies observed in both samples (3 in CNR galaxy subsample and 1 in the comparison sample) is in agreement with the expected one, although the low number of objects make difficult to observe any different tendency. However, the incidence of Sy2 galaxies in the CNR galaxy subsample is significantly higher than that of the comparison sample and the expected value, being 14 against 4 Sy2 galaxies.

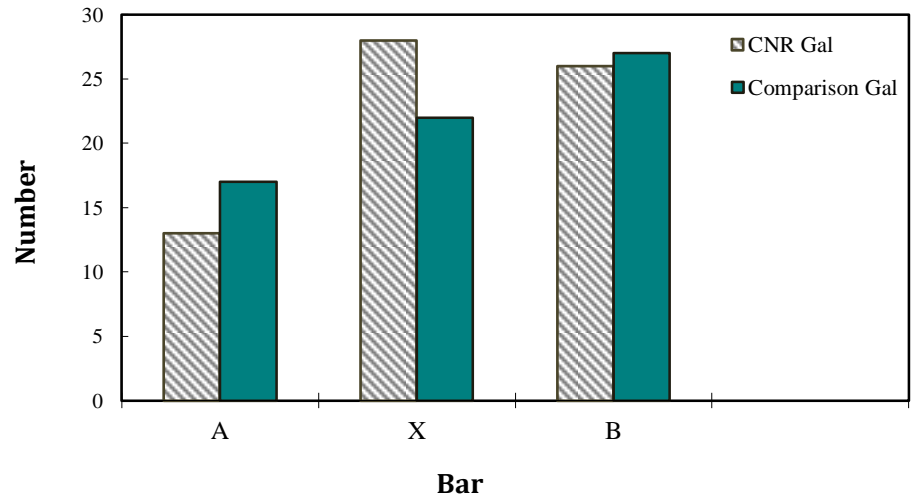

**Figure 5.** Distribution of bar classification for CNR and Comparison galaxy samples.

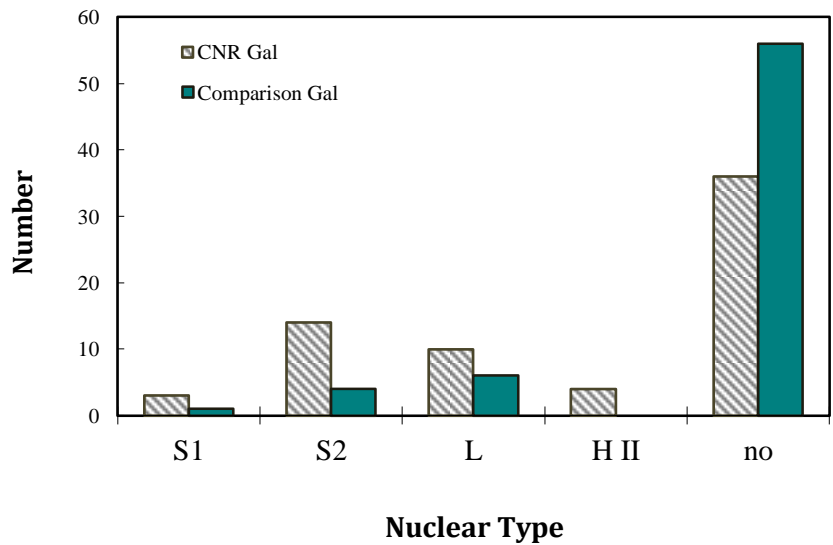

**Figure 6.** Distribution of nuclear classification for CNR and Comparison galaxy samples. . "S1" class includes Sy1 galaxies, Sy1.2, and Sy1.5; "S2" class includes Sy2 galaxies, Sy1.8, Sy1.9 and galaxies classified as Sy1 only in polarized light (Sy1i, Sy1h); "L" class includes LINER galaxies; "HII" class includes galaxies with star formation in the nucleus; "no" includes galaxies with stellar nuclei..

Some properties that could be linked with the CNR formation and evolution are available in [16] catalog. We compare these properties for both samples. For instance, the tidal interaction with nearby galaxies could favor the formation of the CNRs. However, very close encounters could perturb the dynamics of the disk deleting the ILR and the consequent accumulation of material in the CNR. The galaxy local density distribution for both samples is shown in Figure 7. There are no apparent differences in the environment of CNR and comparison galaxies. The content of gas may promote the loss of angular momentum and the consequent feeding of the CNR. Therefore, we compare the amount of neutral gas (M(HI)/$M_{tot}$ and M(HI)/$L_B$) no finding significant differences between both samples. Actually, the M(HI)/LB could present a tendency to be lower for CNR galaxies (Figure 8) although the mean values are indistinguishable. The presence of massive dark halos is invoked for bar stability favoring the generation of resonances. The kinematic width ($W_{20}$) distribution has a peak in 300 Km/s for both samples. Nevertheless, the comparison distribution has a tail toward high velocities which is evident in the median value (307 Km/s for the CNR sample and 410 Km/s for the comparison one). Additionally, we compare the Mass-Luminosity Ratio (Figure 9) observing no apparent differences between samples. The TF relation in B and K bands is consistent with the standard relation of [21] for both samples.

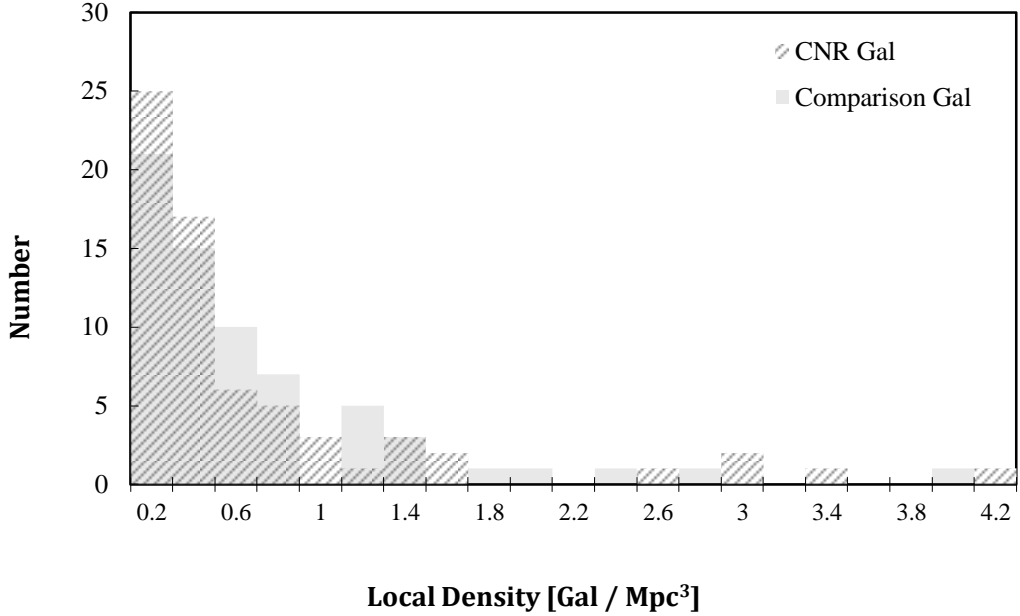

**Figure 7.** Distribution of Local Density of galaxies brighter than -16 B-mag for CNR and Comparison galaxy samples.

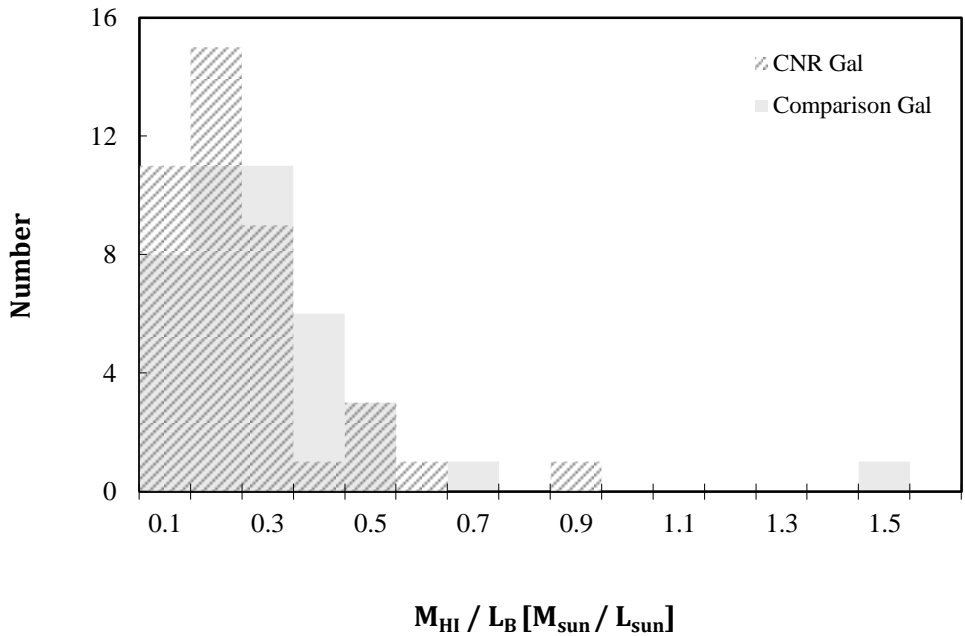

**Figure 8.** Distribution of Mas-Luminosity Ratio in B-band for the neutral gas mass component.

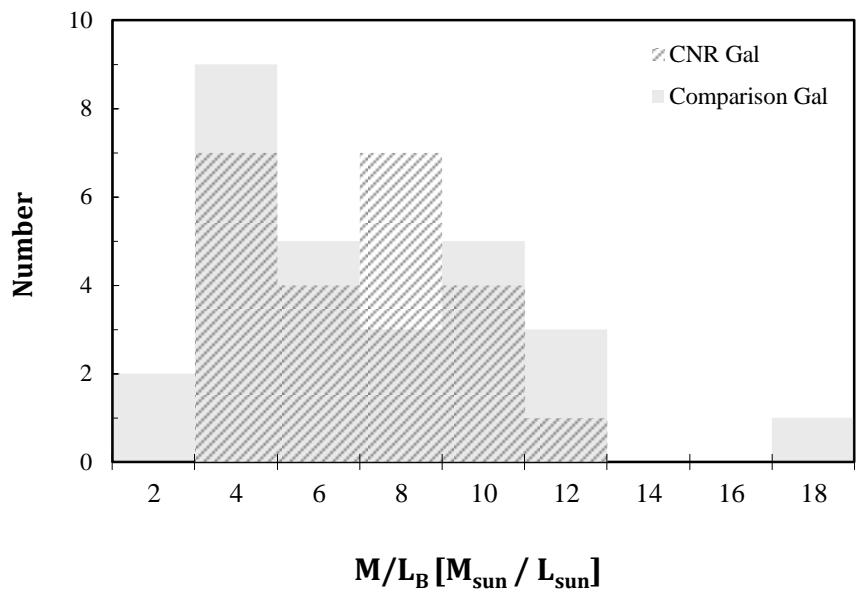

**Figure 9.** Distribution of Mas-Luminosity Ratio in B-band for the galaxy total mass.

## 4. Discussion and Final Remarks

We carry out an exhaustively search of galaxies harboring CNRs in their central region from images, color maps and radial profiles available in the literature, which allowed as to determine the dimensions of the ringed structures. Additionally, we impose some conditions in order to guarantee the circumnuclear nature of the observed ring. Thus, we have set up the most complete sample of 94 CNR galaxies up to 2006. The sample is useful for studying the CNR and host galaxy properties in order to understand the origin and evolution of such structures and the potential connection with the feeding of the central objects.

The CNRs in this sample exhibit a mean diameter of 1.6 Kpc, being the most common value (mode) of 1 Kpc. These results are in agreement with that of [14] who determine a CNR mean diameter of 1.1 Kpc for 20 bared galaxies. The linear diameter of the characteristic CNR is 6% of the host galaxy diameter.

The presence of CNR is more frequent in Sb-Sbc galaxies, and, comparing with the [16] catalog, the CNR galaxy sample present an excess of Sa and Sb galaxies respect to a random sample of 94 spiral galaxies. This morphological type preference could be explain if these ringed structures are favored by evolved disks. The CNR galaxies also prefer weak bars, opposite than expected taking into account that the bar perturbation is invoked as the mechanism of generation of the resonances associated to the CNRs. This result could indicate different evolution time scale for bar and CNR structures.

Dynamical studies propose that the galactic disks can be stabilized by readjusting the mass distribution in the central region, increasing the central rotation velocity which would lead in the ILR development. As a consequence, the density wave propagation is decelerated and the loss of angular momentum is favored [22]. Therefore, the incidence of CNRs, as evidence of the ILR presence, would be greater in galaxies with massive central regions. Considering that the mass of bulge component is related with the mass of the nuclear supermassive black hole [23], a potential connection between the CNRs and the nuclear activity is suspected. Indeed, a larger number of active nuclei are observed in the CNR galaxy sample respect to that expected for a random sample with the same morphological type distribution. Particularly, the main excess is presented in the number of Sy2 active galaxies.

Taking into account the possible selection bias of the sample, we design a comparison galaxies sample with similar observational and physical characteristics. The interlopers (CNR discovered in AGN searches) were excluded and only one source of galaxy properties and nuclear activity classification was used, resulting in 67 CNR galaxy subsample with their respectively comparison galaxy. Although the same tendency of having more weak bars, observed in the full sample, is also insinuated in the subsample, this excess is not conclusive due to the low number of objects. However, the excess of active nuclei respect to the control sample is more robust (17 vs 5). Worth noting that the active galaxies observed in the comparison galaxy sample is in accordance with the expectation obtained from [19] (5 vs 4). Again, the excess of active nuclei is notorious in the Sy2 type, being 14 in the subsample against 4 in the control sample.

In the frame of work of the classical Unified Model for Seyfert galaxies, the Sy1 and Sy2 galaxies are the same phenomenon, which differ in the orientation of the supermassive black hole accretion disk respect to the observer line of sight. Taking into account that the CNRs are more probable observed in galaxies with low inclination, in this sample, we would expect positive selection effect for Sy1 type, unless the accretion disk were not coplanar with the galactic disk in a statistical sense. Some authors propose an evolutionary scenario for Seyfert galaxies in which the active nuclei are powered as the material inflow is increasing (e.g. [1], [2], [25]). In this context, as the gas is led to the central region and the star formation is triggered, the supermassive black hole is fed and the Sy2 stage is activated. As we mentioned, this increasing central mass concentration promotes the CNR formation. And therefore, the coexistence of the CNR and Sy2 phase is expected. However, the increasing inflows toward the nucleus should attenuate the CNR structures in order to when the Sy1 phase is triggered, the CNR is barely observed, as a possible explanation for the lack of excess of Sy1 in the CNR galaxy sample. However, no evolved CNRs are observed in Sy1 galaxies, being 4 CNRs of star formation and 2 dusty CNRs. The [16] catalog allowed us study some other properties in the CNR galaxy subsample and the comparison one. The CNR existence seems to be unrelated with the presence of a bright companion galaxy, whereby the CNR formation would be related to intrinsic galaxy properties or to the merging with satellite dwarf galaxies. The kinematic width ($W_{20}$), M/L ratio and T-F relation do not show evidence for a relation with the CNR presence.

Although the results presented here could be not definitive, the detected trends and their consequences seem strong enough to encourage researchers in the field to perform high quality observations of a large and complete set of galaxies with circumnuclear rings and its matched control sample.

## Acknowledgements

We are in debted to G. Carranza for his support. A preliminary version of this work was presented to the Argentine Astronomical Association in September of 2003. HD thanks the Brazilian institutions Conselho Nacional de Pesquisas (CNPq) and Coordenadora para aperfeiçoamento do Pessoal de Ensino Superior (CAPES). This research is also partially supported by brazilian grants MEGALIT/Millennium.

## References

[1]   Dultzin-Hacyan, D., Krongold, Y., Fuentes-Guridi, I, Marziani, P., 1999, ApJ, 513, L111
[2]   Storchi-Bergmann, T., Gonzalez-Delgado, R.M., Schmitt, H.R., Fernades, R.C., Heckman,T., 2001, ApJ, 559, 147
[3]   Schmitt, H. R., 2001, AAS, 198, 3601
[4]   Virani, S.N., De Robertis, M.M., van Dalfsen, M.L., 2000, AJ, 120, 1739
[5]   Shlosman, I., Frank, J., and Begelman, M.C (1989) Bars within bars - A mechanism for fuelling active galactic nuclei. Nature, **338**, 45-47. http://dx.doi.org/10.1038/338045a0
[6]   Heller, C.H. and Shlosman, I. (1994) Fueling nuclear activity in disk galaxies: Starbursts and monsters. Astrophysical Journal, **424**,84-105. http://dx.doi.org/10.1086/173874
[7]   Maiolino, R., Alonso-Herrero, A., Anders, S., Quillen, A., Rieke, M. J., Rieke, G. H.,and Tacconi-Garman, L. E. (2000) Discovery of a Nuclear Gas Bar Feeding the Active Nucleus in Circinus. Astrophysical Journal, **531**, 219-231. http://dx.doi.org/10.1086/308444
[8]   Combes, F (2000) Bar driven Galaxy Evolution and Time-scales to Feed AGN, in Dynamics of Galaxies: from the Early Universe to the Present, ed. F. Combes, G. A. Mamon, and V. Charmandaris, ASP Conf. Ser. 197, 15.
[9]   Piner, B.G., Stone, J.M., and Teuben, P.J. (1995) Nuclear Rings and Mass Inflow in Hydrodynamic Simulations of Barred Galaxies. Astrophysical Journal, **449**, 508.
[10]  Wada, K. and Habe, A. (1995) Bar-driven fuelling to a galactic central region in a massive gas disc. Monthly Notices of the Royal Astronomical Society, **277**, 433-444. http://dx.doi.org/10.1093/mnras/277.2.433

[11] Storchi-Bergmann, T., Wilson, A.S., and Baldwin, J. A. (1996). Nuclear Rings in Active Galaxies. Astrophysical Journal 460, 252. http://dx.doi.org/10.1086/176966

[12] Díaz, R., Carranza, G.Dottori, H. and Goldes, G. (1999) Kinematics of the Central Regions of NGC 1672. The Astrophysical Journal, **512**, 623-629. http://dx.doi.org/ 10.1086/306781.

[13] Arsenault, R. (1989) The preponderance of bar and ring features in starburst galaxies and active galactic nuclei. Astronomy and Astrophysics **217**, 66-78.

[14] Buta, R. and Crocker, D. A. (1993) Metric characteristics of nuclear rings and related features in spiral galaxies. Astronomical Journal,**105**, 1344-1357. http://dx.doi.org/10.1086/116514

[15] Buta, R.; Crocker, D. A.; Byrd, G. G. (1999) A Hubble Space Telescope Optical and Ground-based Near-Infrared Study of the Giant Nuclear Ring in ESO 565-11. The Astronomical Journal 118, 2071.

[16] Tully, R. B. and Fisher, J. R. (1988) Catalog of Nearby Galaxies, Cambridge University Press.

[17] Véron-Cetty, M.P., and Véron, P. (2003) A catalogue of quasars and active nuclei: 11th edition. Astronomy and Astrophysics, **412**, 399-403. http://dx.doi.org/10.1051/0004-6361:20034225.

[18] Ho, Luis C.; Filippenko, Alexei V.; Sargent, Wallace L. W. (1997) A Search for ``Dwarf'' Seyfert Nuclei. III. Spectroscopic Parameters and Properties of the Host Galaxies. The Astrophysical Journal Supplement Series,**112**,315-390. http://dx.doi.org/ 10.1086/313041

[19] Woltjer,L. (1990) Phenomenology of Active Galactic Nuclei. In: Active Galactic Nuclei, eds. Courvoisier T., Mayor M., Springer-Verlag, Table 5

[20] Veron-Cetty, M. P.; Veron, P. (1989) A Catalogue of quasars and active nuclei. ESO Scientific Report, Garching: European Southern Observatory (ESO), 1989, 4th ed.

[21] Tully, R. Brent; Pierce, Michael J. (2000) Distances to Galaxies from the Correlation between Luminosities and Line Widths. III. Cluster Template and Global Measurement of $H_0$. ApJ 533, 744.

[22] Binney J. & Tremaine S. (1987) Galactic Dynamics, Princeton, NJ, Princeton University Press.

[23] Magorrian, J.; Tremaine, S.; Richstone, D.; Bender, R.; Bower, G.; Dressler, A.; Faber, S. M.; Gebhardt, K.; Green, R.; et al.(1998) The Demography of Massive Dark Objects in Galaxy Centers. AJ 115, 2285.

[24] Koulouridis, E. (201) The dichotomy of Seyfert 2 galaxies: intrinsic differences and evolution. A&A 570, 72.

Table 1: CNR Galaxy Sample

| Galaxy Name | D25 [Kpc] | RCNR [Kpc] | Type | Ref. | laxy Name | D25 [Kpc] | RCNR [Kpc] | Type | Ref. |
|---|---|---|---|---|---|---|---|---|---|
| ESO 138-G1 | 10.8 | 1.25 | S | 1, 2 | NGC 3486 | 13 | 0.95 | H | 18 |
| ESO 437-G33 | 19.4 | 0.6 | H | 3 | NGC 3504 | 19.3 | 0.5 | H | 3, 24 |
| ESO 437-G67 | 22.8 | 0.6 | H | 3 | NGC 3516 | 26.1 | 0.9 | H | 28 |
| ESO 507-G16 | 30.9 | 1.2 | H | 3, 4 | NGC 3593 | 6.4 | 0.2 | H | 10, 44 |
| ESO 565-G11 | 29.3 | 2.6 | H | 3, 5 | NGC 3945 | 38.8 | 0.3 | S | 45 |
| ESO 566-G24 | 21.7 | 1.1 | D | 6 | NGC 3982 | 11.9 | 0.6 | H | 35, 28 |
| IC 1438 | 22.7 | 0.65 | H | 7 | NGC 4151 | 36.2 | 1.9 | D | 46 |
| IC 4214 | 19 | 1.3 | H | 3 ,8 | NGC 4192 | 32.4 | 0.5 | H | 22 |
| MARK 477 | 24.5 | 1.85 | H | 2 | NGC 4303 | 26.2 | 0.6 | H | 28, 47 |
| NGC 0278 | 11 | 0.33 | H | 10, 11 | NGC 4314 | 12.2 | 0.5 | H | 14, 48 |
| NGC 0300 | 6.7 | 0.16 | D | 12 | NGC 4321 | 29.4 | 0.9 | H | 10, 24 |
| NGC 0473 | 20 | 1.3 | H | 13 | NGC 4340 | 13.7 | 0.7 | S | 19 |
| NGC 0521 | 6.2 | 1.95 | S | 14 | NGC 4371 | 21.1 | 0.7 | S | 14, 45 |
| NGC 0613 | 25 | 0.6 | H | 15, 13 | NGC 4526 | 28.9 | 0.6 | D | 49, 50 |
| NGC 936 | 29.6 | 0.8 | S | 16, 17 | NGC 4579 | 26 | 1 | H | 10, 51 |
| NGC 1068 | 33.2 | 1.6 | H | 18, 19 | NGC 4593 | 34.6 | 0.9 | D | 14, 2 |
| NGC 1079 | 24.7 | 0.15 | H | 20 | NGC 4736 | 14.8 | 1 | H | 18 |
| NGC 1097 | 37.7 | 0.55 | H | 21, 15 | NGC 4826 | 8.4 | 0.2 | H | 10, 22 |
| NGC 1241 | 26.4 | 0.71 | H | 22, 23 | NGC 5020 | 41.7 | 0.9 | H | 14 |
| NGC 1300 | 35.7 | 0.5 | H | 10, 24 | NGC 5055 | 24.6 | 0.7 | H | 22 |
| NGC 1317 | 14.8 | 1 | H | 14, 16 | NGC 5194 | 26.5 | 0.6 | H | 16, 52 |
| NGC 1326 | 17.3 | 0.5 | H | 14, 6 | NGC 5236 | 16.1 | 0.15 | D | 53 |
| NGC 1343 | 19 | 1 | H | 25, 13 | NGC 5248 | 37.8 | 0.9 | H | 20, 54 |
| NGC 1365 | 48.4 | 1.7 | H | 16, 26 | NGC 5371 | 44.2 | 0.5 | H | 18 |
| NGC 1386 | 11.8 | 1.1 | H | 1 | NGC 5377 | 33.5 | 0.8 | D | 40, 55 |
| NGC 1433 | 19.6 | 0.5 | H | 27, 20 | NGC 5427 | 25.6 | 1.4 | H | 16, 14 |
| NGC 1512 | 29.1 | 0.27 | H | 16, 20 | NGC 5643 | 28.6 | 0.5 | D | 56 |
| NGC 1530 | 51.3 | 0.7 | H | 16, 28 | NGC 5728 | 25.9 | 0.9 | H | 3, 14 |
| NGC 1543 | 10.2 | 0.7 | S | 19 | NGC 5806 | 21.6 | 0.5 | H | 40, 55 |
| NGC 1566 | 31.3 | 0.9 | H | 29, 30 | NGC 5850 | 38.3 | 1.4 | S | 57, 58 |
| NGC 1672 | 24.1 | 0.5 | H | 31, 2 | NGC 5905 | 55 | 0.6 | H | 14 |
| NGC 1808 | 22.4 | 0.3 | H | 32, 33 | NGC 5945 | 29.1 | 1.3 | H | 4 |
| NGC 1819 | 22.2 | 0.7 | H | 34 | NGC 5953 | 16.4 | 0.9 | H | 13 |
| NGC 2273 | 27.4 | 0.27 | H | 1, 19 | NGC 6221 | 23.8 | 0.9 | H | 59 |
| NGC 2595 | 41.3 | 1 | H | 16 | NGC 6699 | 19.5 | 0.75 | D | 22 |
| NGC 2681 | 14 | 0.6 | D | 16, 35 | NGC 6753 | 28.5 | 1.9 | H | 42, 14 |
| NGC 2763 | 15.8 | 0.45 | H | 36 | NGC 6782 | 30.4 | 1.2 | H | 42, 14 |
| NGC 2903 | 18.4 | 0.4 | H | 37, 24 | NGC 6890 | 13 | 1.1 | H | 2 |
| NGC 2935 | 45.6 | 0.45 | H | 3 | NGC 6951 | 31.7 | 0.5 | H | 14 |
| NGC 2997 | 41.5 | 0.3 | H | 20, 38 | NGC 7187 | 14.3 | 0.8 | S | 14 |
| NGC 3081 | 20.9 | 0.7 | H | 39, 1 | NGC 7217 | 17.8 | 0.9 | D | 14 |
| NGC 3184 | 18.8 | 1.7 | H | 18 | NGC 7469 | 34.4 | 0.7 | H | 60, 2 |
| NGC 3277 | 17.5 | 0.4 | D | 40 | NGC 7552 | 19.9 | 0.5 | H | 61, 62 |
| NGC 3310 | 19.1 | 0.6 | H | 41, 42 | NGC 7570 | 29.1 | 1.1 | H | 13 |
| NGC 3313 | 40.8 | 0.75 | H | 43 | NGC 7582 | 20 | 0.11 | H+D | 9 |
| NGC 3344 | 11.8 | 1 | H | 16, 18 | NGC 7690 | 8.1 | 0.45 | H | 55 |
| NGC 3351 | 16.3 | 0.4 | H | 16, 24 | NGC 7742 | 13 | 1 | H | 14, 16 |

Note: Col. 1: Galaxy Name. Col. 2: Radius of galaxy to the $25^{th}$ mag isophote in B-band in Kpc (NED).
Col. 3: CNR Radius in Kpc. Col. 4: CNR classification: "H" star-forming CNR; "D" Dust CNR, "S" Stellar CNR. Col. 5: Reference for images, color maps or radial profiles where the CNR was identified: (1) Ferruit et al. (2000). (2) Martini et al. (2003). (3) Buta & Crocker (1991). (4) Diaz R. en preparación. (5) Buta et al. (1999). (6) Buta et al. (1998). (7) Crocker et al. (1996). (8) Saraiva (1997. (9) Regan & Mulchaey (1999). (10) Pogge (1989). (11) Knapen et al. (2004). (12) The Messenger 109, 46 (Sep. 2002). (13) Knapen et al. (2006). (14) Wozniak et al. (1995). (15) Hummel et al. (1987). (16) Wray (1988). (17) Erwin & Sparke (2002). (18) Knapen (2005). (19) Erwin (2004). (20) Maoz et al. (1996). (21) Sersic (1958). (22) Böker et al. (1999). (23) Díaz et al. (2003). (24) Pérez-Ramírez et al. (2000). (25) Zwicky & Zwicky (1971). (26) Regan & Vogel (1995). (27) Buta (1986). (28) Knapen et al. (2002). (29) et al. (2004). (30) Sandage & Bedke (1994). (31) Díaz et al. (1999). (32) Kotilainen et al. (1996). (33) Tacconi-Garman et al. (1996). (34) Pogge & Eskridge (1993). (35) Gonzalez Delgado et al. (1997). (36) Böker et al. (2002). (37) Alonso-Herrero et al. (2001). (38) HST F606W image. (39) Storchi-Bergmann et al. (1996). (40) Carollo et al. (2002). (41) Meurer et al. (1995). (42) Windhorst et al. (2002). (43) Buta (1995). (44) Corsini et al. (1998). (45) Erwin & Sparke (1999). (46) Vila-Vilaro et al. (1995). (47) Schinnerer et al. (2002). (48) Benedict et al. (2002). (49) Tomita et al. (2000). (50) King et al. (1995). (51) Boselli et al. (1995). (52) Sakamoto et al. (1999). (53) Elmegreen et al. (1998). (54) Jogee et al. (2002). (55) Carollo et al. (1998). (56) Quillen et al. (1999). (57) Higdon et al. (1998). (58) Lopez Aguerri (1998). (59) Vega Beltran et al. (1998). (60) Kotilainen & Ward (1997). (61) Forbes et al. (1994). (62) Feinstein et al. (1990).

Table 2: CNR Galaxy Properties

| Galaxy Name | T | Bar | Activity | | Galaxy Name | T | Bar | Activity | |
|---|---|---|---|---|---|---|---|---|---|
| ESO 138-G1 | -3 | A | S2 | | NGC 3486 | 5 | X | S2 | |
| ESO 437-G33 | 1 | X | | | NGC 3504 | 2 | X | H2 | |
| ESO 437-G67 | 2 | B | | | NGC 3516 | -2 | B | S1.5 | |
| ESO 507-G16 | 0 | X | | | NGC 3593 | 0 | A | | H |
| ESO 565-G11 | 0 | B | | | NGC 3945 | -2 | B | | L2 |
| ESO 566-G24 | 4 | B | | | NGC 3982 | 4 | X | S1.9 | |
| IC 1438 | 3 | B | | | NGC 4151 | 2 | X | S1.5 | |
| IC 4214 | 2 | X | | | NGC 4192 | 2 | X | S3 | |
| MARK 477 | | | S1h | | NGC 4303 | 4 | X | S2 | |
| NGC 0278 | 3 | X | | H | NGC 4314 | 1 | B | | L2 |
| NGC 0300 | 7 | A | | | NGC 4321 | 4 | X | | T2 |
| NGC 0473 | 0 | X | | | NGC 4340 | -2 | B | | |
| NGC 0521 | 4 | B | | | NGC 4371 | -2 | B | | |
| NGC 0613 | 4 | B | S? | | NGC 4526 | -2 | X | | H |
| NGC 936 | -2 | B | | | NGC 4579 | 3 | X | S3b | |
| NGC 1068 | 3 | A | S1h | | NGC 4593 | 3 | B | S1 | |
| NGC 1079 | 0 | X | | | NGC 4736 | 2 | A | S | |
| NGC 1097 | 3 | B | S3b | | NGC 4826 | 2 | A | S | |
| NGC 1241 | 3 | B | S2 | | NGC 5020 | 4 | X | | |
| NGC 1300 | 4 | B | | | NGC 5055 | 4 | A | | T2 |
| NGC 1317 | 0 | X | | | NGC 5194 | 4 | A | S2 | |
| NGC 1326 | -2 | X | | | NGC 5236 | 5 | X | | |
| NGC 1343 | 3 | X | | | NGC 5248 | 4 | X | | H |
| NGC 1365 | 3 | B | S1.8 | | NGC 5371 | 4 | X | S? | |
| NGC 1386 | -2 | A | S1i | | NGC 5377 | 1 | B | | L2 |
| NGC 1433 | 1 | B | | | NGC 5427 | 5 | A | S2 | |
| NGC 1512 | 1 | B | | | NGC 5643 | 5 | X | S2 | |
| NGC 1530 | 3 | B | | | NGC 5728 | 1 | X | S1.9 | |
| NGC 1543 | -2 | B | | | NGC 5806 | 3 | X | | H |
| NGC 1566 | 4 | X | S1.5 | | NGC 5850 | 3 | B | | L2 |
| NGC 1672 | 3 | B | S | | NGC 5905 | 3 | B | H2 | |
| NGC 1808 | 0 | X | H2 | | NGC 5945 | 2 | B | | |
| NGC 1819 | -2 | B | | | NGC 5953 | -5 | | S2 | |
| NGC 2273 | 1 | B | S1h | | NGC 6221 | 5 | B | S2 | |
| NGC 2595 | 5 | X | | | NGC 6699 | 4 | X | | |
| NGC 2681 | 0 | X | S3b | | NGC 6753 | 3 | A | | |
| NGC 2763 | 6 | B | | | NGC 6782 | 1 | X | | |
| NGC 2903 | 4 | X | | H | NGC 6890 | 3 | A | S1.9 | |
| NGC 2935 | 4 | X | | | NGC 6951 | 4 | X | S2 | |
| NGC 2997 | 5 | X | | | NGC 7187 | -1 | X | | |
| NGC 3081 | 1 | X | S1h | | NGC 7217 | 2 | A | S3 | |
| NGC 3184 | 6 | X | | H | NGC 7469 | 1 | X | S1.5 | |
| NGC 3277 | 2 | A | | | NGC 7552 | 2 | B | H2 | |
| NGC 3310 | 4 | X | H2 | | NGC 7570 | 1 | B | | |
| NGC 3313 | 2 | B | | | NGC 7582 | 2 | B | S1i | |
| NGC 3344 | 4 | X | | H | NGC 7690 | 3 | A | | |
| NGC 3351 | 3 | B | | H | NGC 7742 | 3 | A | | T2/L2 |

Note: Col. 1: Galaxy Name. Col. 2: Morphological Type from Tully (1988: E=-5; S0=-2; Sa=1; Sb=3; Sc=7). Col. 3: Bar Intensity Classification ("B": Strong; "X" Weak; "A" Absent). Col. 4: Nuclear Activity Classification: (*left*) Veron-Cetty & Veron (2006) nuclear classification ("H2" nuclear star-formation; "S1", "S1.5", "S1.8", "S1.9", and "S2" Seyfert nuclei; "S1h" Sy2 nuclei with broad emission lines only in polarized light; "S3" LINER/Transition); (*right*) Ho et al. (1997) nuclear classification ("H" nuclear star-formation; "L2" LINER; "T2" Transition).

Table 3: Properties of CNR Galaxy Subsample

| Name | Comp.Gal. | T | B | i | $D_{25}$ | $2R_{25}$ | $M_B$ | $M_K$ | D | rho | $W_R$ | $\log(M_H)$ | $\log(M_T)$ | $M_H/M_T$ | $M_H/L_B$ | $M_T/L_B$ |
|------|-----------|---|---|---|------|------|-------|-------|------|------|------|-------|-------|-------|-------|-------|
| IC 1438 | UGC 9837 | 3 | B | 25 | 2.3 | 22.7 | -19.45 | -23.34 | 33.8 | 0.17 | | | | | | |
| NGC 278 | NGC 4369 | 3 | XP | 0 | 2.7 | 11 | -19.62 | -22.54 | 11.8 | 0.11 | | | 9.02 | | | 0.1 |
| NGC 300 | UGC 5666 | 7 | A | 46 | 20.2 | 6.7 | -16.88 | -19.02 | 1.2 | 0.45 | 200 | 8.9 | 9.89 | 0.102 | 0.9 | 8.83 |
| NGC 473 | NGC 357 | 0 | X | 54 | 2.5 | 20 | -19.77 | -22.81 | 29.8 | 0.08 | | | | | | |
| NGC 613 | NGC 4254 | 4 | B | 37 | 5 | 25 | -20.53 | -24.19 | 17.5 | 0.07 | 577 | 9.49 | 11.38 | 0.013 | 0.12 | 9.54 |
| NGC 936 | NGC 2217 | -2 | B | 40 | 6.2 | 29.6 | -20.08 | -24.23 | 16.9 | 0.24 | | | | | | |
| NGC 1068 | UGC 2953 | 3 | A | 29 | 8 | 33.2 | -21.39 | -25.00 | 14.4 | 0.34 | | | 9.22 | | 0.03 | |
| NGC 1079 | NGC 1415 | 0 | XP | 49 | 5.4 | 24.7 | -18.83 | -22.80 | 16.9 | 0.2 | | | | | | |
| NGC 1097 | NGC 3992 | 3 | B | 57 | 9.8 | 37.7 | -20.79 | -24.56 | 14.5 | 0.13 | 436 | 9.87 | 11.32 | 0.036 | 0.23 | 6.45 |
| NGC 1241 | NGC 7412 | 3 | B | 52 | 3.6 | 26.4 | -19.83 | -23.47 | 26.6 | 0.11 | | | | | | |
| NGC 1300 | NGC 4123 | 4 | B | 43 | 6.8 | 35.7 | -20.42 | -23.81 | 18.8 | 0.71 | 373 | 9.55 | 11.16 | 0.025 | 0.15 | 6.28 |
| NGC 1317 | NGC 3729 | 0 | X | 38 | 3.1 | 14.8 | -19.2 | -23.41 | 16.9 | 1.3 | | | | | | |
| NGC 1326 | NGC 3941 | -2 | B | 53 | 3.8 | 17.3 | -19.84 | -23.69 | 16.9 | 1.36 | 305 | 9.26 | 10.67 | 0.039 | 0.13 | 3.47 |
| NGC 1365 | NGC 4569 | 3 | B | 63 | 11.2 | 48.4 | -21.26 | -24.77 | 16.9 | 1.58 | 410 | 10.04 | 11.37 | 0.046 | 0.22 | 4.75 |
| NGC 1433 | NGC 4665 | 1 | B | 27 | 5.9 | 19.6 | -19.67 | -23.26 | 11.6 | 0.55 | | | | | | |
| NGC 1512 | NGC 1055 | 1 | B | 77 | 13.4 | 29.1 | -19.02 | -22.40 | 9.5 | 0.37 | 239 | 9.53 | 10.68 | 0.069 | 0.53 | 7.65 |
| NGC 1530 | NGC 1169 | 3 | B | 60 | 4.8 | 51.3 | -21.32 | -24.53 | 36.6 | 0.2 | 334 | 10.09 | 11.22 | 0.074 | 0.23 | 3.16 |
| NGC 1543 | NGC 4503 | -2 | B | 65 | 3 | 10.2 | -19.07 | -23.19 | 13.4 | 0.95 | | | | | | |
| NGC 1566 | NGC 1398 | 4 | X | 40 | 8.4 | 31.3 | -20.45 | -23.75 | 13.4 | 0.92 | 305 | | 10.93 | | | 3.59 |
| NGC 1672 | NGC 2775 | 3 | B | 37 | 5.9 | 24.1 | -19.84 | -23.79 | 14.5 | 0.64 | | | | | | |
| NGC 1808 | NGC 1371 | 0 | X | 50 | 7.6 | 22.4 | -19.52 | -23.79 | 10.8 | 0.3 | | | 9.26 | | 0.18 | |
| NGC 2273 | NGC 5678 | 1 | B | 50 | 3.4 | 27.4 | -20.76 | -23.79 | 28.4 | 0.19 | 430 | 9.46 | 11.17 | 0.02 | 0.09 | 4.68 |
| NGC 2681 | NGC 4457 | 0 | X | 0 | 3.6 | 14 | -19.59 | -23.19 | 13.3 | 0.2 | | | 9.12 | | 0.12 | |
| NGC 2763 | NGC 6412 | 6 | | 0 | 2 | 15.8 | -19.53 | -22.49 | 25.7 | 0.22 | | | 9.46 | | 0.29 | |
| NGC 2903 | NGC 3521 | 4 | X | 66 | 11.6 | 18.4 | -19.85 | -22.96 | 6.3 | 0.12 | 391 | 9.25 | 10.91 | 0.022 | 0.13 | 6.02 |
| NGC 2935 | NGC 2207 | 4 | X | 49 | 5.4 | 45.6 | -21.74 | -24.12 | 30.6 | 0.3 | 359 | 10.01 | 11.23 | 0.06 | 0.13 | 2.2 |
| NGC 2997 | NGC 3359 | 5 | X | 53 | 10.2 | 41.5 | -20.74 | -24.29 | 13.8 | 0.28 | 301 | 10.06 | 11.04 | 0.105 | 0.37 | 3.55 |
| NGC 3081 | NGC 5691 | 1 | X | 35 | 2.2 | 20.9 | -20.05 | -23.65 | 32.5 | 0.25 | | | | | | |
| NGC 3184 | NGC 2835 | 6 | X | 26 | 7.5 | 18.8 | -19.34 | -22.44 | 8.7 | 0.17 | | | 9.36 | | 0.27 | |

| Nombre | Comp.Gal. | T | B | i | $D_{25}$ | $2R_{25}$ | $M_B$ | $M_K$ | D | r | $W_R$ | $\log(M_H)$ | $\log(M_T)$ | $M_H/M_T$ | $M_H/L_B$ | $M_T/L_B$ |
|---|---|---|---|---|---|---|---|---|---|---|---|---|---|---|---|---|
| NGC 3277 | NGC 718 | 2 | A | 26 | 2.4 | 17.5 | -19.51 | -23.06 | 25 | 0.35 | | | | | | |
| NGC 3310 | NGC 3887 | 4 | XP | 25 | 3.6 | 19.1 | -20.09 | -22.76 | 18.7 | 0.27 | | 9.69 | | | 0.29 | |
| NGC 3344 | NGC 1637 | 4 | X | 23 | 6.7 | 11.8 | -18.47 | -21.49 | 6.1 | 0.19 | | 9.25 | | | 0.47 | |
| NGC 3351 | NGC 4274 | 3 | B | 56 | 7.5 | 16.3 | -19.26 | -22.87 | 8.1 | 0.54 | 302 | 8.92 | 10.64 | 0.019 | 0.1 | 5.48 |
| NGC 3486 | NGC 1249 | 5 | X | 47 | 6.4 | 13 | -18.61 | -21.35 | 7.4 | 0.27 | 285 | 9.27 | 10.48 | 0.061 | 0.43 | 7.06 |
| NGC 3504 | NGC 4750 | 2 | X | 35 | 2.6 | 19.3 | -20.38 | -23.85 | 26.5 | 0.25 | | | | | | |
| NGC 3516 | NGC 5087 | -2 | B | 34 | 2.3 | 26.1 | -20.59 | -24.44 | 38.9 | 0.19 | | 9.38 | | | 0.09 | |
| NGC 3593 | NGC 4460 | 0 | A | 69 | 4.8 | 6.4 | -17.01 | -21.28 | 5.5 | 0.19 | 232 | 8.02 | 10 | 0.01 | 0.11 | 10.12 |
| NGC 3945 | NGC 5365 | -2 | B | 55 | 6.5 | 38.8 | -20.3 | -24.23 | 22.5 | 0.5 | | | | | | |
| NGC 3982 | UGC 3190 | 4 | X | 26 | 2.4 | 11.9 | -19.45 | -22.30 | 17 | 0.72 | | 9.21 | | | 0.18 | |
| NGC 4192 | NGC 4216 | 2 | X | 83 | 8.7 | 32.4 | -21.08 | -24.24 | 16.8 | 1.44 | 429 | 9.75 | 11.24 | 0.033 | 0.13 | 4.11 |
| NGC 4303 | NGC 2442 | 4 | X | 17 | 5.9 | 26.2 | -20.71 | -24.43 | 15.2 | 1.06 | | 9.66 | | | 0.15 | |
| NGC 4314 | NGC 1947 | 1 | B | 15 | 4.2 | 12.2 | -18.65 | -22.48 | 9.7 | 1.25 | | | | | | |
| NGC 4321 | NGC 488 | 4 | X | 37 | 6.1 | 29.4 | -21.13 | -24.54 | 16.8 | 2.95 | 444 | 9.7 | 11.23 | 0.03 | 0.11 | 3.82 |
| NGC 4340 | NGC 1387 | -2 | B | 20 | 2.8 | 13.7 | -19.17 | -22.81 | 16.8 | 2.89 | | | | | | |
| NGC 4371 | NGC 2685 | -2 | B | 66 | 5 | 21.1 | -19.3 | -23.41 | 16.8 | 4.09 | | | | | | |
| NGC 4526 | NGC 4438 | -2 | X | 74 | 7.4 | 28.9 | -20.55 | -24.66 | 16.8 | 2.45 | | 9.31 | | | 0.08 | |
| NGC 4579 | NGC 3631 | 3 | X | 36 | 5.4 | 26 | -20.67 | -24.64 | 16.8 | 3.26 | 605 | 9.36 | 11.44 | 0.008 | 0.08 | 9.57 |
| NGC 4593 | NGC 3223 | 3 | B | 58 | 3.3 | 34.6 | -21.58 | -24.99 | 39.5 | 0.21 | | | | | | |
| NGC 4736 | NGC 3368 | 2 | A | 33 | 12.2 | 14.8 | -19.37 | -23.06 | 4.3 | 0.42 | 373 | 8.48 | 10.78 | 0.005 | 0.03 | 6.86 |
| NGC 4826 | NGC 2683 | 2 | A | 66 | 8 | 8.4 | -19.15 | -22.73 | 4.1 | 0.2 | | 8.32 | | | 0.03 | |
| NGC 5055 | NGC 4559 | 4 | A | 55 | 13 | 24.6 | -20.14 | -23.68 | 7.2 | 0.4 | 447 | 9.65 | 11.15 | 0.032 | 0.26 | 8.05 |
| NGC 5194 | NGC 4258 | 4 | AP | 64 | 13.6 | 26.5 | -20.75 | -23.93 | 7.7 | 0.33 | | 9.46 | | | 0.09 | |
| NGC 5236 | NGC 4051 | 5 | X | 24 | 11.5 | 16.1 | -20.31 | -23.74 | 4.7 | 0.18 | | 10.01 | | | 0.5 | |
| NGC 5248 | NGC 1232 | 4 | X | 45 | 6.1 | 37.8 | -21.07 | -24.53 | 22.7 | 0.3 | 361 | 10.05 | 11.15 | 0.079 | 0.27 | 3.43 |
| NGC 5371 | NGC 2417 | 4 | X | 41 | 4.2 | 44.2 | -21.57 | -25.28 | 37.8 | 0.74 | 585 | 10.12 | 11.64 | 0.03 | 0.2 | 6.63 |
| NGC 5377 | NGC 5448 | 1 | B | 58 | 4.2 | 33.5 | -20.69 | -24.10 | 31 | 0.19 | | | | | | |
| NGC 5427 | NGC 3367 | 5 | AP | 17 | 2.3 | 25.6 | -20.94 | -24.31 | 38.1 | 0.19 | | | | | | |
| NGC 5728 | NGC 4699 | 1 | X | 65 | 2.3 | 25.9 | -21.67 | -24.96 | 42.2 | 0.18 | | | | | | |
| NGC 5806 | NGC 5633 | 3 | X | 63 | 2.9 | 21.6 | -20.36 | -23.82 | 28.5 | 0.85 | | | | | | |
| NGC 5850 | NGC 5375 | 3 | B | 28 | 4.6 | 38.3 | -20.69 | -24.17 | 28.5 | 0.65 | | 9.62 | | | 0.14 | |

| Nombre | Comp.Gal. | T | B | i | $D_{25}$ | $2R_{25}$ | $M_B$ | $M_K$ | D | r | $W_R$ | $\log(M_H)$ | $\log(M_T)$ | $M_H/M_T$ | $M_H/L_B$ | $M_T/L_B$ |
|---|---|---|---|---|---|---|---|---|---|---|---|---|---|---|---|---|
| NGC 5953 | NGC 5017 | -5 | | 43 | 1.8 | 16.4 | -19.59 | -23.43 | 33 | 0.3 | | | | | | |
| NGC 6221 | NGC 3672 | 5 | B | 55 | 3.9 | 23.8 | -21.04 | -24.32 | 19.4 | 0.14 | | | | | | |
| NGC 6951 | NGC 5350 | 4 | X | 28 | 3.7 | 31.7 | -20.73 | -24.69 | 24.1 | 0.08 | | 9.68 | | | 0.16 | |
| NGC 7217 | NGC 4691 | 2 | A | 32 | 3.6 | 17.8 | -20.38 | -24.17 | 16 | 0.15 | | 8.81 | | | 0.03 | |
| NGC 7552 | NGC 3507 | 2 | B | 31 | 3.5 | 19.9 | -20.14 | -23.91 | 19.5 | 0.52 | | 9.57 | | | 0.21 | |
| NGC 7690 | NGC 6239 | 3 | A | 73 | 2.1 | 8.1 | -18.87 | -21.58 | 16.4 | 0.09 | | | | | | |
| NGC 7742 | NGC 4412 | 3 | A | 28 | 2 | 13 | -19.65 | -23.10 | 22.2 | 0.1 | | | | | | |

Note: Col. 1: CNR Galaxy Name. Col. 2: Comparison Galaxy Name. Col. 3: Morphological Type (E=-5; S0=-2; Sa=1; Sb=3; Sc=7). Col. 4: Bar Classification. Col 5: galaxy inclination [*deg*]. Col. 6: Diameter of galaxy to the $25^{th}$ mag isophote in B-band [*arcmin*].Col. 7: Diameter of galaxy to the $25^{th}$ mag isophote in B-band corrected by projection and extinction [*Kpc*]. Col. 8: Absolute Magnitude in B-band. Col. 9: Absolute Magnitude in K-band. Col. 10: Galaxy Distance (h=0.75) [*Mpc*]. Col. 11: Local Density of Galaxies brighter than $M_B$=-16 [*Galaxy/Mpc³*]. Col. 12: KinematicWidth $W^i_R = 2\ V_{max}$ [*Km/s*]. Col. 13: Mass of neutral gas [$M_\odot$]. Col. 14: Total Mass [$M_\odot$]. Col. 15: Fraction of nuclear gas. Col. 16: Mass-Luminosity ratio for neutral gas in B-band [$M_\odot/L_\odot$]. Col. 17: Total Mass-Luminosity ratio in B-band [$M_\odot/L_\odot$].